\documentclass[aps,prb,twocolumn,showpacs]{revtex4}

\usepackage{times}
\usepackage{color}
\usepackage{graphicx}
\usepackage{dcolumn}
\usepackage{amssymb}
\usepackage{amsmath}
\usepackage{amsfonts}
\usepackage{docs}%
\usepackage{bm}%
\usepackage[colorlinks=true,linkcolor=blue,pagecolor=blue,filecolor=blue,menucolor=blue,urlcolor=blue,citecolor=blue,anchorcolor=blue]{hyperref}%

\newcommand{\ie}{\textit{i.e. }}

\begin{document}

\title{Tunable Supercurrent at the Charge Neutrality Point via Strained Graphene Junctions}

\author{Mohammad Alidoust }
\author{Jacob Linder}
\affiliation{Department of Physics, Norwegian University of Science
and Technology, N-7491 Trondheim, Norway}

\date{\today}

\begin{abstract}
We theoretically calculate the charge-supercurrent through a
ballistic graphene junction where superconductivity is induced via
the proximity-effect. Both monolayer and bilayer graphene are
considered, including the possibility of strain in the systems. We
demonstrate that the supercurrent at the charge neutrality point can
be tuned efficiently by means of mechanical strain. Remarkably, the
supercurrent is enhanced or suppressed relative to the non-strained
case depending on the direction of this strain. We also calculate
the Fano factor in the normal-state of the system and show how its
behavior varies depending on the direction of strain.
\end{abstract}


\maketitle

\section{Introduction}
Since the discovery of graphene \cite{cite:novoselov1}, many
interesting phenomena have been predicted in the context of quantum
transport in this material \cite{cite:beenakker1, cite:Titov1,
cite:Tworzydlo, cite:Snyman, cite:Ossipov}. It has been demonstrated
in several theoretical and experimental works that the conductivity
of graphene monolayer junctions at zero doping level (the charge
neutrality point \textit{aka} Dirac point) displays a minimum value
in the short junction regime
\cite{cite:novoselov1,cite:zhang,cite:Tworzydlo,
cite:Titov1,cite:Akhmerov}. The physical reason for the minimum
conductivity in this regime is the existence of evanescent modes
that can transport current over a finite length $L$ of the system. A
unique aspect of graphene is that these evanescent modes are
predicted to generate pseudo-diffusive characteristics of the
quantum transport properties even in ballistic graphene samples
\cite{cite:Tworzydlo}.

Bilayer graphene is a basic carbon
structure and has attracted considerable attention recently. This system
consists involved two coupled graphene layers with prominent characteristics such as 
pseudospin and variable chirality \cite{cite:choi2}. The chirality of the massless Dirac fermions in
monolayer graphene is locked to the momentum direction and consequently lies in the plane of the sheet. In bilayer graphene,
however, massive Dirac fermions with perpendicular chirality to the
sheet plane may occur. Bilayer graphene features a band-structure similar to a 
semiconductor with parabolic bands with a tunable charge excitation gap
\cite{cite:choi1,zhang1,ostinga,cite:castro,kuzmenko,li,maik}. A gapless graphene bilayer is more stable compared to its gapped equivalence and thus occurs naturally. Bilayer graphene has also shown anomalous phenomena such as
half-integer quantum Hall effect, minimum conductivity at zero energy and
$2\pi$ Berry phase. 

Very recently, the response of graphene to strain has been
intensively examined. Several unusual effects have been unveiled,
including the generation of very high pseudo-magnetic fields of
order 300 T \cite{cite:levy_science_10}. An interesting question in
this context relates to if strain imposed on graphene, whether it be
mechanical or thermal in origin, can be used to control its
transport properties \cite{cite:peres_rmp_10}. This issue is also
motivated by the well-known fact that strain imposed on
silicon-based devices can enhance their functionality. A mechanical
deformation of a graphene sheet will invariably generate scattering
centers which effectively influences the hopping amplitude, and thus
suggests that the transport of Dirac fermions should respond to the
presence of strain. It is known that strain may change the physical properties of
nanotubes drastically \cite{yang2, tombler, minot,
cite:choi1,cite:choi2}. The strain can be induced in graphene via several routes, including mechanically
\cite{mohi,ferralis,kim,huang,teague}.

Based on this idea, we address in this paper a novel class of the
Josephson graphene junctions with the capability to sustain tuneable
charge-transport at the Dirac point by means of \textit{mechanically
induced strain}. To demonstrate this, we first solve the Bogoliubov
de-Gennes equations both for a strained monolayer and bilayer
graphene-based superconductor$\mid$normal$\mid$superconductor
(S$\mid$N$\mid$S) junction with heavily doped S regions, as is
experimentally relevant. We then derive explicit analytical
expressions for the Andreev-levels and use these to obtain the
phase-dependent supercurrent $I(\phi)$ in the short-junction regime
\cite{cite:beenakker2}. Both the critical current $I_c$ and the
$I_cR_N$ product is investigated for a range of doping levels in the
N region, including the charge neutrality point. Above, $R_N$ is the
normal state resistance. Finally, we calculate the Fano factor $F$
in the normal (non-superconducting) state and show how this is
influenced by the presence of mechanical strain in the system.

\begin{figure}[t!]
\includegraphics[width=6.5cm,height=5.5cm]{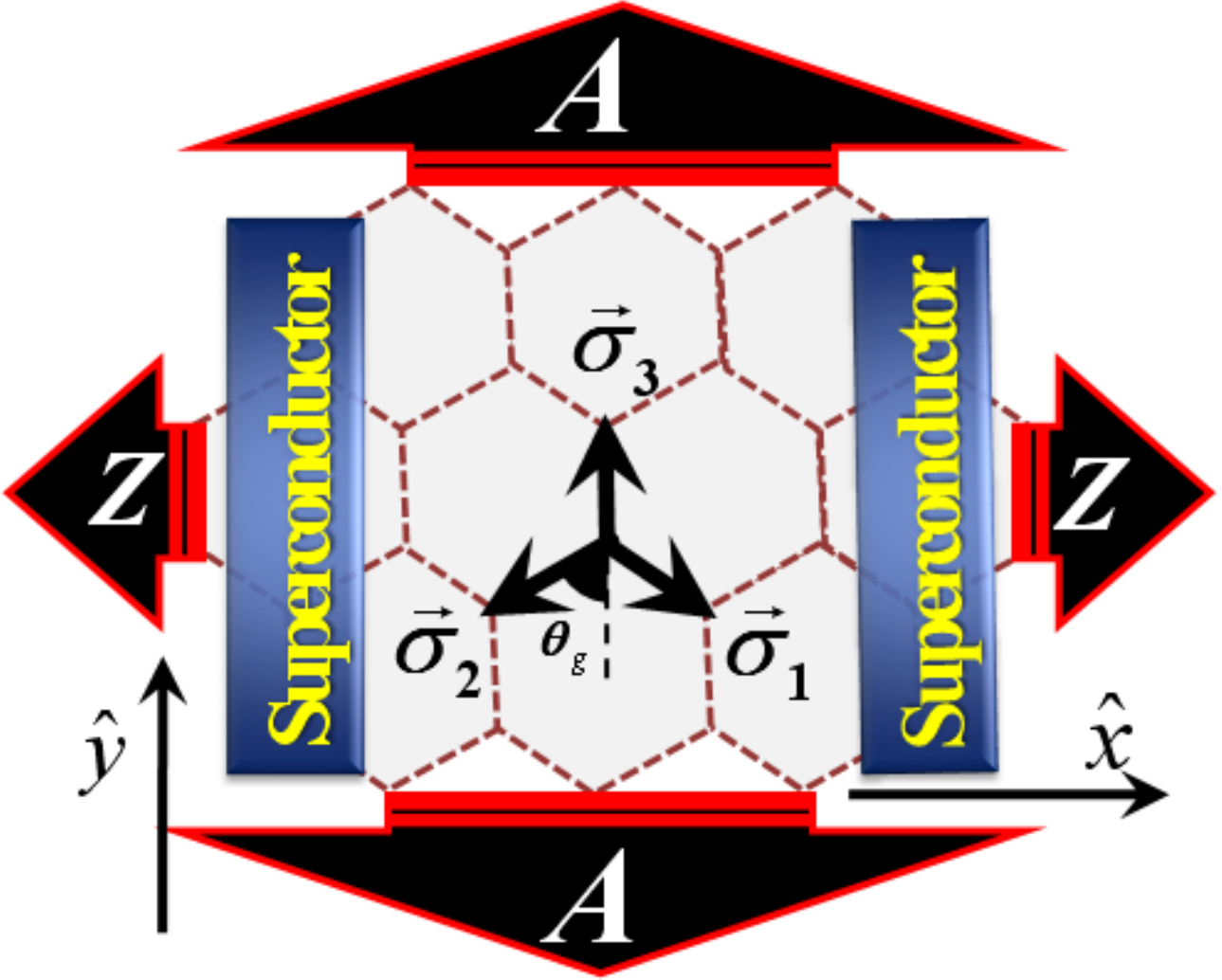}
\caption{\label{fig:model} The schematical setup of a strained
graphene Josephson junction. The two superconducting electrode
interfaces are located at $x=0, L$. There is also a controllable
gate voltage for tuning the concentration of carriers (not shown).
$Z$ and $A$ stand for \textit{zig-zag} and \textit{armchair} strains
while $\sigma_{i}$ represent displacement vectors of the three
nearest neighbors C atoms in the strained graphene. $\theta_{g}$ is
the strained angle between C-C junctions which can be either larger
or smaller than the non-strained value $60^{\circ}$.}
\end{figure}

To describe strained graphene, we adopt the model used in
Ref.\onlinecite{cite:Pereira} for monolayer graphene and also
consider a similar model for strained graphene bilayer junctions.
Our findings show that for a \textit{zig-zag} ($Z$)-strain (see
Fig.\ref{fig:model}) the transmission probability of evanescent
modes near the charge neutrality point is suppressed, which
influences both the conductivity $\sigma$ and the supercurrent.
However, when the strain is applied along the \textit{armchair}
($A$) direction instead, the transmission probability is enhanced
and correspondingly influences charge-transport in the system. These
results point towards new perspectives within tunable quantum
transport by means of induced strain in a graphene mono- or bilayer.
This finding might also be of relevance in the field of spintronics
(valleytronics), since the strain also affects the pseudo-spin of
the chiral fermions in graphene \cite{cite:choi1,cite:choi2}.

The paper is organized as follows: In Sec. \ref{section:theory}, we
present our theoretical approach and derive a general expression for the 
normal-state transition probabilities describing both strained monolayer and
bilayer junctions. In this section, we also discuss an experimental setup for
detecting our predictions. In Sec. \ref{section_resultsanddiscussion}, the 
Andreev subgap bound state energies for both strained monolayer and
bilayer S$\mid$N$\mid$S Josephson junctions are obtained and the subsequent results are 
discussed. We finally conclude our findings in Sec.
\ref{sec:conclusion}.

\section{Theoretical Approach and Model}\label{section:theory}
The two graphene systems considered in this paper (monolayer and bilayer) are
modelled via the following Hamiltonians
\cite{cite:beenakker3,cite:mccann1,cite:nilsson,cite:Snyman,cite:castro}
($\mathcal{S}$ and $\mathcal{B}$ stand for single- and bilayer
graphene under strain, respectively):
        \begin{equation}\label{eq:1L_Hamiltonian}
        \mathcal{H}_{\pm}^{\mathcal{S}}=v_xp_x\sigma_x\pm
v_yp_y\sigma_y +U
        \end{equation}
        \begin{equation}\label{eq:2L_Hamiltonian}
\mathcal{H}^{\mathcal{B}}=\left(
          \begin{array}{cccc}
            U & \pi & 0 & 0 \\
            \pi^{\dag} & U & t_{\bot} & 0 \\
            0 & t_{\bot} & U & \pi^{\dag} \\
            0 & 0 & \pi & U \\
          \end{array}
        \right)
        \end{equation}
where $\pi$$=$$v_xp_x$$+$$iv_yp_y$ and $v_{x,y}$ are the Fermion
velocities in the $\hat{x}, \hat{y}$-directions while $\sigma_{x,y}$
are Pauli matrices. Here, $U$ represents an external gate potential.
The $\pm$ signs refer to the $\mathbf{K}$ and $\mathbf{K'}$ valleys
of monolayer graphene. We here consider an $A_2B_1$ stack for the bilayer
graphene sheet and use a tight-binding model in which the massive
chiral Dirac frmions are governed by Eq. (\ref{eq:2L_Hamiltonian}).
The resulting Hamiltonian becomes similar to a gapless semiconductor Hamiltonian with
parabolic electron and hole bands touching when
$t$$\rightarrow$$\infty$. In order to model applied strain to the
system, we adopt the model of Ref. \onlinecite{cite:Pereira} and
expand the tight-binding model band structure with arbitrary hopping
energies $t_{1,2,3}$ \ie $\epsilon=\pm\mid
\sum_{i=1}^{3}t_ie^{-i\overrightarrow{k}\cdot\overrightarrow{\sigma_{i}}}\mid$
around the Dirac point,
$\mathbf{K}_D=(\cos^{-1}(-1/2\eta)/\sqrt{3}a_x,0)$
\cite{cite:choi1,cite:soodchomshom}. As shown in Fig.
\ref{fig:model}, $\overrightarrow{\sigma_{i}}$ are displacement
vectors between three nearest C atoms (see Ref.
\onlinecite{cite:paper_sigma_i}). We assume $t_{1,2}=t_{\vdash}$ and
$t_3=t$ in our calculations and set $\eta$ as the ratio
$t_{\vdash}/t$. This assumption generates asymmetric Fermion
velocities along the different directions. These velocities are
given by $v_x=2t_{\vdash}a_x\sin(\cos^{-1}(-1/2\eta))/\hbar$ and
$v_y=3t a_y/2\hbar$ \cite{cite:choi1,cite:choi2,cite:soodchomshom}.
The next-nearest neighbor hopping (n.n.n.) in graphene can cause the
Dirac cone to be tilted, this effect vanishes under the influence of
strain of order $\sim$ 20\% \cite{cite:choi1}. In this regime, the
generalized Weyl-Hamiltonian used here gives very good agreement
with \textit{ab initio} calculations. We note that it is also
possible to model strained graphene by including a fictious
gauge-potential $\mathbf{A}$ \cite{peres_rmp_10} . Motivated by the
results of Ref. \onlinecite{cite:Snyman} for bilayer graphene
junctions, we here adopt the same model as the strained monolayer
for the strained bilayer whereas the interlayer hopping $t_{\bot}$
is left intact as the strain is applied in-plane. We
\textit{emphasize} that in this paper the same model for
clean doped bilayer graphene region is adopted as Ref.
\onlinecite{cite:Snyman} and consequently the trigonal warping
effects may be neglected.

In both cases, we consider an S$\mid$N$\mid$S junction with $s$-wave
superconducting electrodes. Previous works have been considered the Josephson effect in
non-strained monolayer graphene \cite{cite:Titov1,cite:cuevas,blackschaffer_prb_08}. We
assume that the Fermi wavelengths satisfy $\lambda_N \gg\lambda_S$,
corresponding to heavily doped S regions. In this regime, we may
ignore interface details and consider the following $x$-dependent
superconducting order parameter as depicted in Fig.\ref{fig:model}
(see Ref. \onlinecite{cite:halterman});
\begin{eqnarray}\label{eq:Delta}
\Delta(x)=\left\{\begin{array}{cc}
  \Delta(T)e^{i\phi_{l}} & x<0 \\
  0 & 0<x<L \\
  \Delta(T)e^{i\phi_{r}} & x>L \\
\end{array}\right..
\end{eqnarray}
By substituting the Hamiltonians Eq.(\ref{eq:1L_Hamiltonian}) and
Eq.(\ref{eq:2L_Hamiltonian}) into the Bogoliubov-de Gennes equation
\begin{equation}\label{eq:BdG}
\left(\begin{array}{cc}
 \mathcal{H}-\mu & \Delta \\
  \Delta^{*}& \mu-\mathcal{H}
\end{array}\right)
\end{equation}
where $\mu$ is the chemical potential, we find the following
energy-momentum dispersion relations:
\begin{equation}\label{eq:2L_dispersion}
\left\{\begin{array}{c}
  \varepsilon^{\mathcal{B}}=\left[\Delta^{2}+\left\{\mu\pm\frac{t_{\bot}}{2}\pm\frac{1}{2}\sqrt{t_{\bot}^2+4\hbar^2|k|^{2}\nu^2}\right\}^{2}\right]^{\frac{1}{2}} \\
    \varepsilon^{\mathcal{S}}=\left[\Delta^{2}+\left\{\mu\pm\hbar |k|\nu\right\}^{2}\right]^{\frac{1}{2}} \\
    \nu^{2}=v_{x}^{2}\cos^{2}\theta+v_{y}^{2}\sin^{2}\theta
\end{array}\right.
\end{equation}
Above, $\theta$ is the angle of incidence of the particles. The
eigenfunctions of Eq.(\ref{eq:BdG}) for the strained monolayer and
bilayer systems within the normal region are found to be:
\begin{align}
^{\mathcal{S}}\left\{\begin{array}{c}
  \Psi_{e^{\pm}}^{N}=(\pm a_{e^{\pm}}^{N},1,\mathbf{0}^{\mathcal{S}})^{T}e^{\pm i\hbar
k_{e}^{N}x} \\
  \Psi_{h^{\pm}}^{N}=(\mathbf{0}^{\mathcal{S}},\mp a_{h^{\pm}}^{N},1)^{T}e^{\pm i\hbar
k_{h}^{N}x}
\end{array}\right.
\end{align}
\begin{align}
^{\mathcal{B}}\left\{\begin{array}{c}
  \Psi_{e^{\pm}}^{N}=(-1,\mp a_{e^{\pm}}^{N},\pm a_{e^{\pm}}^{N},1,\mathbf{0}^{\mathcal{B}})^{T}e^{\pm i\hbar
k_{e}^{N} x} \\
  \Psi_{h^{\pm}}^{N}=(\mathbf{0}^{\mathcal{B}},-1,\mp a_{h^{\pm}}^{N},\pm a_{h^{\pm}}^{N},1)^{T}e^{\pm i\hbar
k_{h}^{N} x}
\end{array}\right.
\end{align}
where $\mathbf{0}^{\mathcal{S}}$ and $\mathbf{0}^{\mathcal{B}}$
represent $1\times 4$- and $1\times 2$-spinors with only zeroes as
entries, while $e$ and $h$ stand for electron and hole particles.
The $\pm$ sign refers to right and left going particles. The
coefficients $a_{e^{\pm}}^{N}$ and $a_{h^{\pm}}^{N}$ are defined as
follows:
\begin{align}\label{eq:1L_ae}
^{\mathcal{S}}\left\{\begin{array}{c}
 a_{e^{\pm}}^{N}=\frac{\mu+\varepsilon}{\hbar k_{e}^{N}( v_x\cos\theta\pm iv_{y}\sin\theta)} \\
  a_{h^{\pm}}^{N}=\frac{\mu-\varepsilon}{\hbar
  k_{h}^{N}(
  v_x\cos\theta_A\pm iv_{y}\sin\theta_A)} \\
  \hbar k_{e(h)}^{N}=\frac{\mu(\pm)\varepsilon}{\nu}
\end{array}\right.
\end{align}
\begin{align}\label{eq:2L_ae}
^{\mathcal{B}}\left\{\begin{array}{c}
 a_{e^{\pm}}^{N}=\frac{\mu+\varepsilon}{\hbar k_{e}^{N}( v_x\cos\theta\pm iv_{y}\sin\theta)} \\
  a_{h^{\pm}}^{N}=\frac{\mu-\varepsilon}{\hbar
  k_{h}^{N}(
  v_x\cos\theta_A\pm iv_{y}\sin\theta_A)} \\
  \hbar
  k_{e(h)}^{N}=\sqrt{\frac{t(\mu(\pm)\varepsilon)}{\nu^{2}}}
\end{array}\right.
\end{align}

For the bilayer system, we consider the hopping value between the
graphene layers to be smaller than the doping level of
superconducting regions, still much larger than the superconducting
gap that is $\mu^s\gg t\gg\varepsilon_F,\varepsilon$
 \cite{cite:ludwig}. The former
assumption assures ignoring the contact details in the S$\mid$N
interfaces while the latter not only helps to simplify theoretical
approach but also warranties realistic approximations in our
analytical calculations. Within the normal region in which
superconducting order parameter $\Delta$$=$$0$, the Eq.
(\ref{eq:BdG}) leads to uncoupled equations.  In this paper, we focus
on the low-energy regime. One then finds the
following parabolic dispersion relation for electrons and holes in
the normal bilayer region:
\begin{equation}\label{2L_dispersion_limiting}
    \varepsilon^{\mathfrak{B}}=\left| \varepsilon_F\pm\left(\frac{(\hbar v
    |k|)^{2}}{t}\right)\right|.
\end{equation}
Due to translational symmetry in the transverse direction, $k_{y}$
and $\varepsilon$ are both conserved upon reflections at the
interfaces located at $x=0, L$. Accordingly, the dispersion
relations and the following equation assure both energy and momentum
conservation of particles in the $y$-direction upon Andreev
electron-hole conversion,
\begin{eqnarray}\label{conserv}
\nonumber k_{e}^{N}\sin\theta=k_{h}^{N}\sin\theta_A=q_n=2n\pi/W\\
k_{e}^{N}\sin\theta=k_{e,h}^{S}\sin\theta_{e,h}^{S}=q_n=2n\pi/W.
\end{eqnarray}

The total spinors in the three regions thus read:
\begin{eqnarray}\label{eq:totalspinors}
\nonumber \Psi^{N}&=& e^{i
\hbar q_{n}y}(a_1\Psi_{e^{+}}^{N}+a_2\Psi_{e^{-}}^{N}+b_1\Psi_{h^{+}}^{N}+b_2\Psi_{h^{-}}^{N})\\
\Psi_{r}^{S}&=& e^{i\hbar
q_ny}(t_{e}^{r}\Psi_{e^{+}}^{S}(\phi_r)+t_{h}^{r}\Psi_{h^{+}}^{S}(\phi_r))\\
\nonumber \Psi_{l}^{S}&=& e^{i\hbar
q_ny}(t_{e}^{l}\Psi_{e^{-}}^{S}(\phi_l)+t_{h}^{l}\Psi_{h^{-}}^{S}(\phi_l)).
\end{eqnarray}
In superconductor spinors, we define the following relation for
superconducting coherent factors (see Appendix):
\begin{equation}\label{beta}
    \beta=\left\{\begin{array}{cc}
      \cos^{-1}(\varepsilon/\Delta) & \varepsilon<\Delta \\
      -i\cosh^{-1}(\varepsilon/\Delta) & \varepsilon>\Delta \\
    \end{array}\right.
\end{equation}
the definition helps to simplifying our notation. The spinors in the
superconducting regions carry $S$ superscript, whereas $r$ and $l$
stand for right and left superconducting regions. The
superconducting phases in each region are assumed to be $\phi_r$ and
$\phi_l$, while $t_{h}$ and $t_{e}$ are the scattering amplitudes of
hole- and electron-like quasiparticles. Matching the total
wavefunctions at each of the interfaces, \ie
\begin{equation}\label{eq:matching}
\Psi_{l}^{S}\mid_{x=0}=\Psi^{N}\mid_{x=0}\;\;\;\text{and}\;\;\;\Psi_{r}^{S}\mid_{x=L}=\Psi^{N}\mid_{x=L}
\end{equation}
leads a quantization relation between superconducting phase
difference $\phi=\phi_r-\phi_l$ and the quasiparticle excitation
energy $\varepsilon$. The boundary conditions lead to a 8$\times$8
matrix $\emph{M}$ for transmission and reflection coefficients which
is presented in the Appendix for monolayer case \cite{cite:linder1}.
$det (\emph{M})=0$ generates a non-trivial relation between
$\varepsilon$ and $\phi$ as follow;
\begin{equation}\label{andreev_bound}
    \digamma_1+\digamma_2\sin2\beta+\digamma_3\cos2\beta=0,
\end{equation}
\begin{eqnarray}
 \nonumber&&\digamma_1=-(a_{e^{-}}^{N}+a_{e^{+}}^{N}) (a_{h^{-}}^{N}+a_{h^{+}}^{N}) \cos \phi\\\nonumber&&+ \sin (k_{h}^{N} L)\sin (k_{e}^{N} L)((-a_{e^{+}}^{N}
   a_{h^{-}}^{N}-a_{h^{+}}^{N} a_{h^{-}}^{N}+a_{e^{+}}^{N} a_{h^{+}}^{N}\\\nonumber&&-a_{e^{-}}^{N} (-a_{h^{-}}^{N} a_{h^{+}}^{N}
   a_{e^{+}}^{N}+a_{e^{+}}^{N}-a_{h^{-}}^{N}+a_{h^{+}}^{N})+1)
 \\
 \nonumber&&\digamma_2=\sin (k_{e}^{N} L)(a_{e^{-}}^{N} a_{e^{+}}^{N}+1)
   (a_{h^{-}}^{N}+a_{h^{+}}^{N}) \cos (k_{h}^{N} L) \\\nonumber&&+(a_{e^{-}}^{N}+a_{e^{+}}^{N}) \cos
   (k_{e}^{N}L)(a_{h^{-}}^{N}
   a_{h^{+}}^{N}+1) \sin (k_{h}^{N} L)\\
  \nonumber&&\digamma_3= -\sin (k_{h}^{N} L)\sin (k_{e}^{N} L)(a_{e^{-}}^{N} a_{e^{+}}^{N}+1) (a_{h^{-}}^{N}
   a_{h^{+}}^{N}+1) \\\nonumber&&+(a_{e^{-}}^{N}+a_{e^{+}}^{N}) \cos
   (k_{e}^{N}L) (a_{h^{-}}^{N}+a_{h^{+}}^{N}) \cos (k_{h}^{N}L)
\end{eqnarray}
We here employ the most relevant experimentally approximation \ie
the "\textit{short-junction}" regime in which $\Delta L/\hbar v\ll
1$. Within this regime, $\digamma_1$, $\digamma_2$ and $\digamma_3$
are reduced to the following expressions;
\begin{eqnarray}
 \nonumber&&\digamma_1=-(a_{e^{-}}^{N}+a_{e^{+}}^{N})^{2} \cos \phi+ \sin^{2} (k L)
 \\\nonumber&&\times((a_{e^{+}}^{N})^{2}+
   a_{e^{-}}^{N} (-(a_{e^{+}}^{N})^{2} a_{e^{-}}^{N}
   +a_{e^{-}}^{N})-1)
 \\
&&\digamma_2=0\\
  \nonumber&&\digamma_3= \sin^{2} (kL)(a_{e^{-}}^{N} a_{e^{+}}^{N}+1)^{2}+(a_{e^{-}}^{N}+a_{e^{+}}^{N})^{2}
  \cos^{2}
   (kL).
\end{eqnarray}
In this case, using the definition of $\beta$ and
Eq.(\ref{andreev_bound}) the single Andreev bound state is obtained
vs $\digamma_1$ and $\digamma_3$
\begin{eqnarray}\label{eq:boundstate}
  &&\nonumber\frac{\varepsilon}{\Delta}=\sqrt{\frac{1}{2}\left(1-\frac{\digamma_1}{\digamma_3}\right)}\;\;\text{and
then}\\
  &&\nonumber\varepsilon_{n}(\phi)=\Delta\sqrt{1-\tau_{n} \sin^{2}\phi/2}
\end{eqnarray}
in which $\tau_n$ is transmission probability for the normal
graphene region between two strongly doped electrodes (either
superconductor or normal). After some calculations we reach at an
expression for $\tau_n$ valid for both strained monolayer and
bilayer systems \ie
\begin{equation}\label{eq:normalpropability}
    \tau_{n}=\frac{(a_{e^{+}}^{N}+a_{e^{-}}^{N})^{2}}{(a_{e^{+}}^{N}+a_{e^{-}}^{N})^{2}\cos^{2}(k_{n}L)+(a_{e^{+}}^{N}a_{e^{-}}^{N}+1)^{2}\sin^{2}(k_{n}L)}.
\end{equation}
We utilize the general $\tau_n$ for investigating the transport
properties of the strained monolayer and bilayer junctions in the
next section.

The contribution of the
Andreev bound-state spectrum to the supercurrent is given by
\cite{cite:Titov1,cite:beenakker2}:
\begin{equation}\label{eq:Josephson_current}
    I(\phi)=\frac{e\Delta^2}{\hbar}\sum_{n=0}^{\infty}\tau_{n}\sin\phi/\varepsilon_n(\phi).
\end{equation}
For a wide graphene junction, $W$$\gg$$L$, the boundary conditions
(zig-zag and armchair) at the edges $y$$=$$W/2$ and $-W/2$ are
irrelevant and we here assume smooth boundaries for the two edges.
In this regime, we may replace the summation over quantized modes
with an integration: $\sum_{n}$$\rightarrow$$W/\pi$$\int$$dq_{n}$.
Now we proceed in the following sections to study the 
supercurrent using Eqs. (\ref{eq:1L_ae}), (\ref{eq:2L_ae}) and
(\ref{eq:Josephson_current}) for monolayer and bilayer Josephson
junctions in particular at the charge neutrality point \ie
$\mu\rightarrow 0$.

\section{Supercurrent, Fano Factor and Andreev Bound States in
Strained Graphene Monolayer/Bilayer S$\mid$N$\mid$S
Junction}\label{section_resultsanddiscussion}

\begin{figure}[t!]
\includegraphics[width=8.5cm,height=5.2cm]{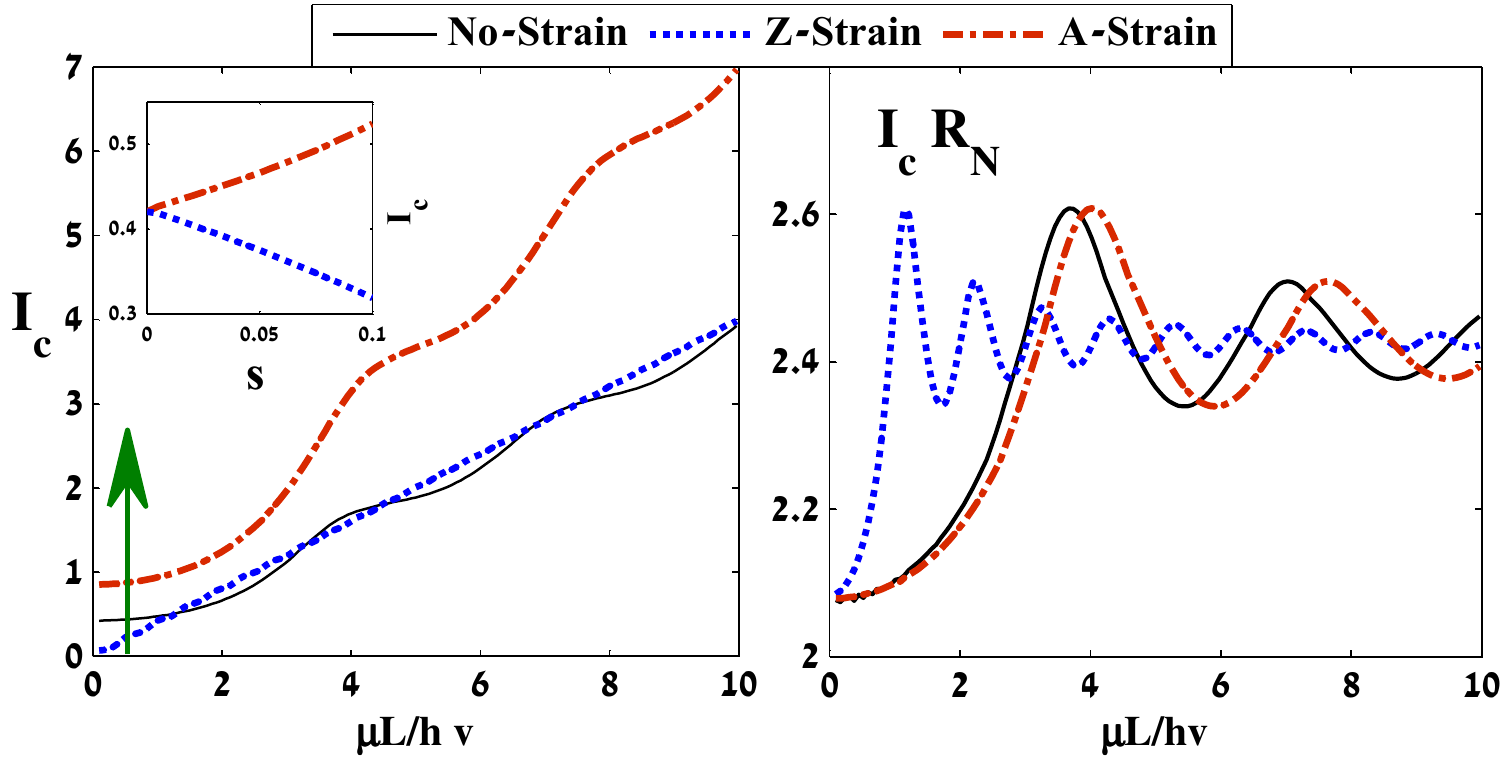}
\caption{\label{fig:monolayer} The critical current $I_c$ (left
panel) and its product with the normal-state resistance $I_cR_N$
(right panel) as a function of $\mu L/\hbar v$ for a monolayer
system.  The solid line pertains to a non-strained junction. For
$Z$-tension $t_{\vdash}=0.56t_0,\; t=1.1t_0$ while for $A$-strain
$t_{\vdash}=0.95t_0,\; t=0.5t_0$. The arrow indicates how the
Josephson current may be enhanced by means of the applied direction
of tension to the system. The inset panel shows critical
supercurrent as a function of strain.}
\end{figure}

By inserting Eq.(\ref{eq:1L_ae}) for the monolayer system into
Eq.(\ref{eq:normalpropability}), the following expressions are
obtained:
\begin{equation}\label{eq:1l_tau}
    \nonumber\tau_{n}^{\mathcal{S}}=\frac{[v_{x}\cos\theta_{n}^{\mathcal{S}}]^{2}}{[v_{x}\cos\theta_{n}^{\mathcal{S}}\cos(k_{n}^{\mathcal{S}}L)]^{2}+[{\nu_{n}^{\mathcal{S}}}\sin(k_{n}^{\mathcal{S}}L)]^{2}}\\
\end{equation}
\begin{equation}
   \nonumber
    {k_{n}^{\mathcal{S}}}=\sqrt{\frac{{\mu}^{2}}{(\hbar
    {\nu_{n}^{\mathcal{S}}})^{2}}-q_{n}^{2}},\;\theta_{n}^{\mathcal{S}}=\text{atan}\left(\frac{\hbar^2v_{x}^{2}q_{n}^{2}}{{\mu}^{2}-\hbar^2v_{y}^{2}q_{n}^{2}}\right)^{\frac{1}{2}}.
\end{equation}
In the case of a bilayer-system where Eq.(\ref{eq:2L_ae}) holds, the
corresponding normal-state transmission probability takes the from
\begin{equation}\label{eq:2l_tau}
    \nonumber\tau_{n}^{\mathcal{B}}=\frac{\mu[2v_{x}\cos\theta_{n}^{\mathcal{B}}]^{2}/\nu_{n}^{\mathcal{B}}}{\frac{\mu[2v_{x}\cos\theta_{n}^{\mathcal{B}}\cos(k_{n}^{\mathcal{B}}L)]^{2}}{\nu_{n}^{\mathcal{B}}}+t_{\bot}[(\mu/t_{\bot}+1)\sin(k_{n}^{\mathcal{B}}L)]^{2}}\\
\end{equation}
\begin{equation}
    \nonumber k_{n}^{\mathcal{B}}=\sqrt{\frac{\mu t_{\bot}}{(\hbar
    {\nu_{n}^{\mathcal{B}}})^{2}}-q_{n}^{2}},\;\theta_{n}^{\mathcal{B}}=\text{atan}\left(\frac{\hbar^2v_{x}^{2}q_{n}^{2}}{\mu
    t_{\bot}-\hbar^2v_{y}^{2}q_{n}^{2}}\right)^{\frac{1}{2}}.
\end{equation}
\begin{figure}[t!]
\includegraphics[width=8.5cm,height=5.2cm]{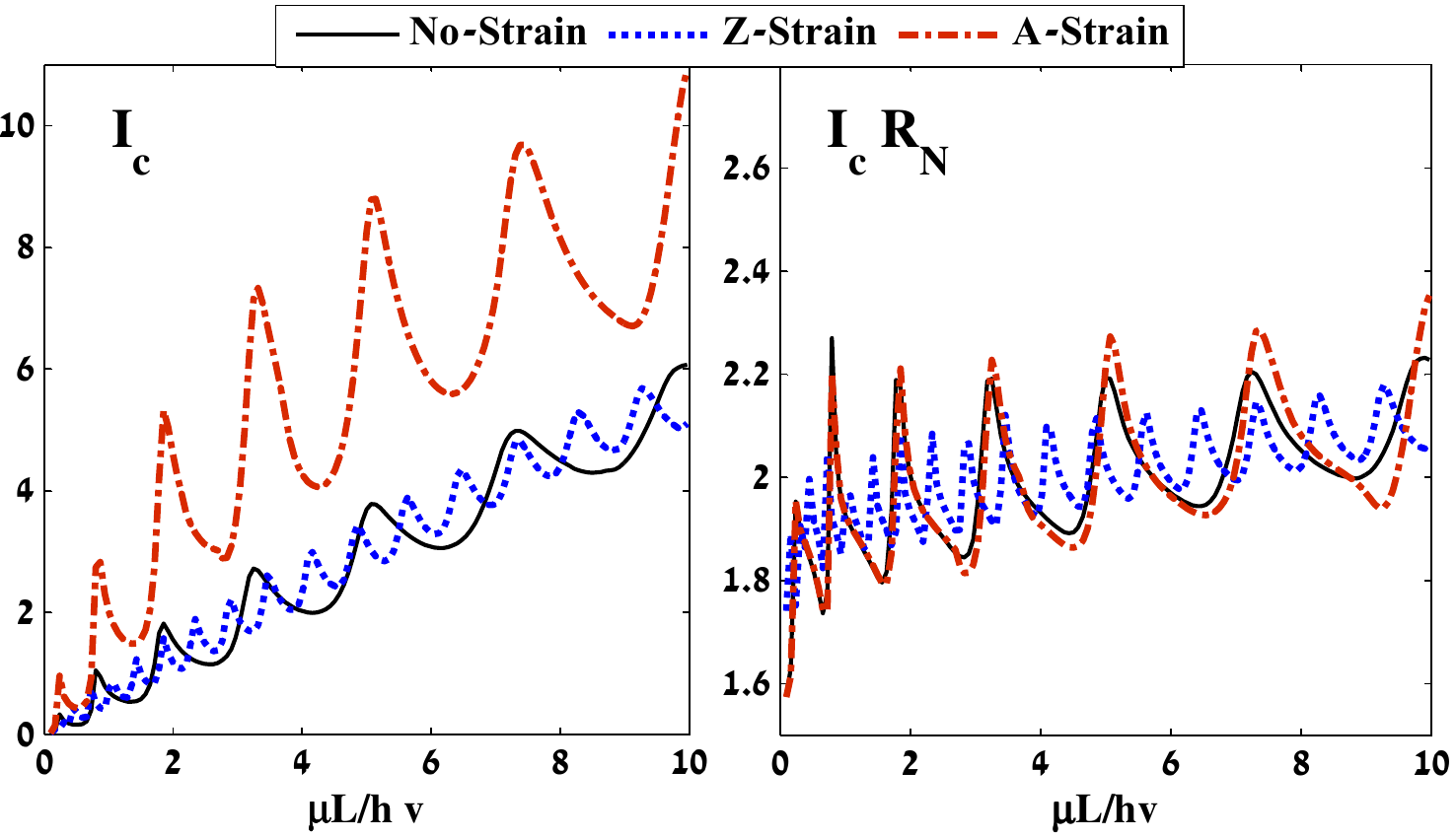}
\caption{\label{fig:bilayer} The critical current $I_c$ (left panel)
and its product with the normal-state resistance $I_cR_N$ (right
panel) as a function of $\mu L/\hbar v$ for a strongly coupled
bilayer system. The solid line pertains to a non-strained junction.
Here, the same values as for the monolayer system have been used for
the tensions, \ie in the case of $Z$-tension $t_{\vdash}=0.56t_0,\;
t=1.1t_0$ while for $A$-tension $t_{\vdash}=0.95t_0,\; t=0.5t_0$ for
$s=0.2$.}
\end{figure}
The behavior of $I_c$ and $I_cR_N$ for strained monolayer
graphene-based Josephson junction vs $\mu L/\hbar v$ is shown in
Fig.\ref{fig:monolayer}. The critical current against strain is
shown in the inset panel (see Ref.\onlinecite{cite:Ribeiro}). For
$A$-strain, we assume $t_{\vdash}=0.95t_0$, $t=0.5t_0$ ($t_0=2.7$ eV
for non-strained graphene) and for $Z$-strain $t_{\vdash}=0.56t_0$,
$t=1.1t_0$ which follows when $s=0.2$ \cite{cite:Pereira}. As seen,
the critical current becomes nearly zero at the charge neutrality
point for $Z$-strain, whereas the current is enhanced compared to
the non-strained case when the strain is applied along the $A$
direction. In effect, the $Z$-strain induces a very small
phase-dependent contribution to the Andreev-bound states and the
supercurrent vanishes. This suggests a remarkable fact: \textit{the
supercurrent can be efficiently tuned by means of both the magnitude
and the direction of the strain imposed on the system, even at the
Dirac point}. We have also considered the same model of strain for a
bilayer graphene S$\mid$N$\mid$S junction and plot $I_c$ and
$I_cR_N$ as a function of $\mu L/\hbar v$ in Fig.\ref{fig:bilayer}.
In this case, it is seen that the critical current tends toward zero
as one approaches the Dirac point because of the assumption
$t_{\bot}\gg\varepsilon, \Delta$ which influences the evanescent
modes \cite{cite:paper_delta_F}. The obtained normal-state
transmission probability $\tau_n^\mathcal{B}$ is proportional to
$\mu$ and hence tends toward zero as one approaches the charge
neutrality point. In order to understand these results, one has to
consider two facts: \textit{i)} all transport modes $n$ in the
system become evanescent $(k_n = i q_n)$ at the Dirac point and
\textit{ii)} have a transmission probability through the junction
given by $\tau_n$. In the $Z$- and $A$-strain cases, $\tau_n$
decays, respectively, faster and slower than the non-strained
graphene as a function of $q_n$. In turn, this dictates the
magnitude of the contribution of transverse modes to the electron
transmission and thus to the discrete Andreev bound state spectrum
for the $A$- and $Z$-strains with respect to the non-strained
system.
\begin{figure}[t!]
\includegraphics[width=8.5cm,height=5.2cm]{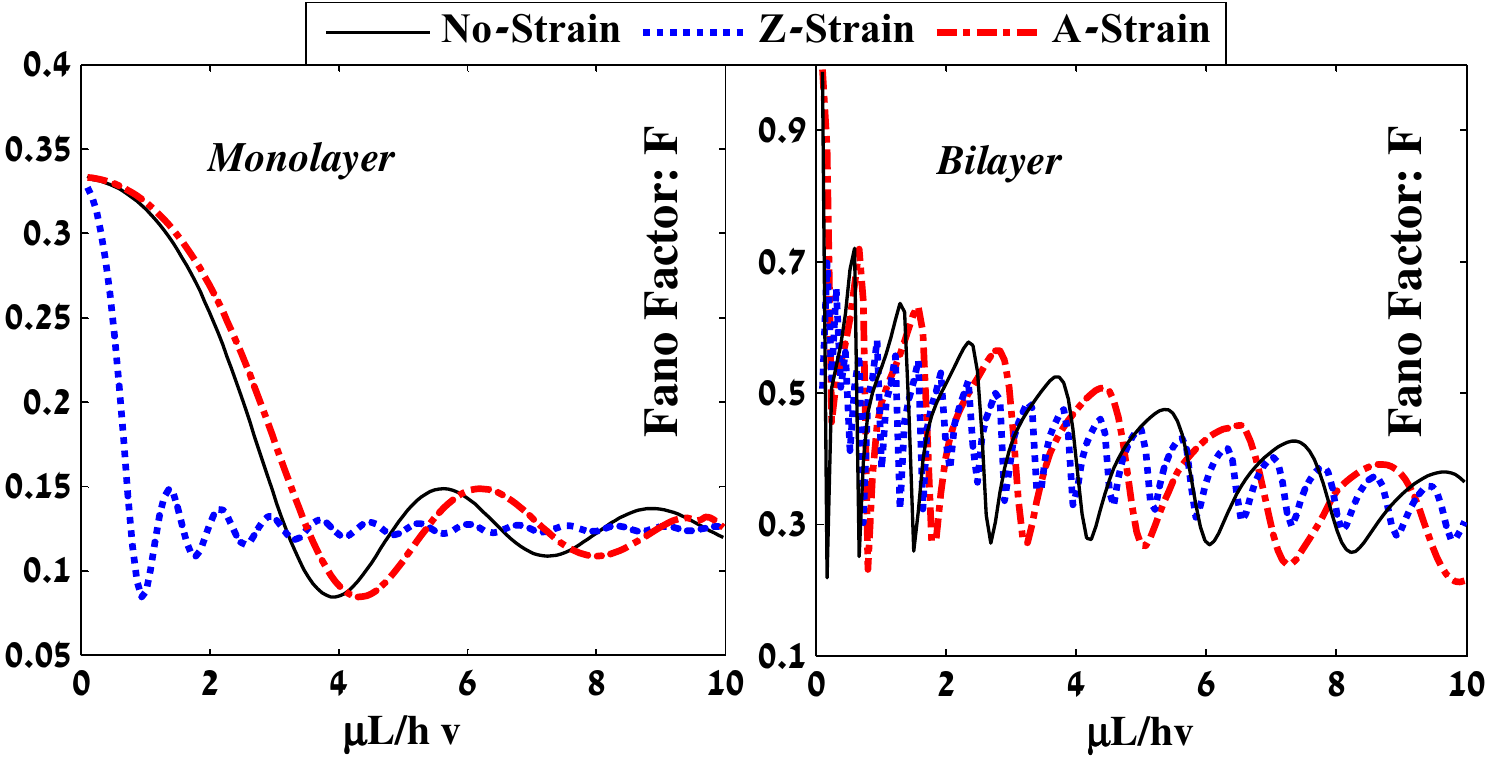}
\caption{\label{fig:fanofactor} The Fano factor $F$ for both
monolayer and bilayer systems with a weakly doped region sandwiched
between two heavily doped sides (the normal-state of the mentioned
S$\mid$N$\mid$S junction) as a function of $\mu L/\hbar v$. In the
scenarios with strain, the same parameters as in
Fig.\ref{fig:monolayer} and Fig.\ref{fig:bilayer} have been used.}
\end{figure}
We have also calculated the Fano factor (the ratio of noise power
and mean current) via the normal transmission probability $\tau_n$,
defined as $F = \sum^{\infty}_{0} \tau_n (1-\tau_n) /
\sum^{\infty}_{0} \tau_n$ \cite{cite:Tworzydlo}. The results for
both mono- and bilayer graphene with and without strain are shown in
Fig.\ref{fig:fanofactor}. In the non-strained case, we reproduce
previous results for monolayer \cite{cite:Tworzydlo} and bilayer
\cite{cite:Snyman} junctions where a weakly doped middle region is
sandwiched between two heavily doped regions
\cite{cite:Ossipov,cite:Katsnelson}. The scenario with strain has
not been considered up to now, and inspection of
Fig.\ref{fig:fanofactor} reveals that the strain influences how $F$
evolves with the doping-level in the middle region. More
specifically, in the $Z$-strain case the contribution of the
transversal modes is suppressed and therefore the $F$ goes towards
saturation faster than the $A$- and non-strained regimes as the
doping-degree $\mu$ is increased. The system under tension, however,
sustains still the universal value of $F=1/3$ at the Dirac point
just as the non-strained monolayer system or diffusive normal metal
\cite{cite:Tworzydlo,cite:Snyman}. We note that the influence of
trigonal warping may be neglected in the monolayer case when the
impurity-potential is weak (ballistic regime) \cite{tworz_prb_08}.
For the bilayer case, the trigonal warping becomes influential in
the low-energy regime $|\varepsilon| < 0.5\gamma_1(v_3/v)^2$ where a
relevant estimate for the parameters is $\gamma_1=0.39$ eV and
$v_3/v = 0.1 \ll 1$  \cite{cite:mccann1}. This yields
$|\varepsilon|<2$ meV. However, the influence of strain in the
considered bilayer model in this paper becomes most evident at
higher doping levels as seen from Fig.\ref{fig:bilayer} where the
trigonal warping effects can be neglected.

\section{Conclusion}\label{sec:conclusion}
In conclusion, we have proposed a novel class of ballistic graphene
monolayer/bilayer-based Josephson junctions with mechanical strain.
We have derived a general analytical normal transition probability
valid for both strained monolayer and bilayer graphene systems. We
have demonstrated that the direction of the applied strain to the
system near the charge neutrality point can be used to efficiently
tune the magnitude of the supercurrent in such a system. In
addition, we have considered the Fano factor $F$ in the normal-state
of this junction and how it is influenced by strain in the system.
In this case, we also find that the direction of the strain is
influential with respect to how $F$ depends on the doping-level of
the graphene sheet. We believe that these results point towards new
perspectives within tunable quantum transport by means of
mechanically induced strain. Interesting phenomena may be 
expected to arise out of the coexistence of proximity induced ferromagnetism and
superconductivity in a strained graphene junctions
\cite{cite:zou}.

\appendix
\section{\label{appendix}Andreev subgap states}
In this appendix, we present more details of our analytical approach
used to find the general normal transition probability
in Sec. \ref{section:theory}. We here also examine our
analytical expressions for non-strained monolayer case where $s$ is
equal to zero. In the graphene monolayer superconducting regions, the
right- and left-going quasiparticles are described via the following
spinors
\begin{eqnarray}\label{eq:totalspinors}
\left\{\begin{array}{c}
  \Psi_{e^{\pm}}^{S}=(\pm a_{e^{\pm}}^{S}e^{i\beta},e^{i\beta},\pm a_{e^{\pm}}^{S},1)^Te^{\pm ik_{e}^{S}L} \\
  \Psi_{h^{\pm}}^{S}=(\mp a_{e^{\mp}}^{S},1,\mp
a_{e^{\mp}}^{S}e^{i\beta},e^{i\beta})^Te^{\mp ik_{h}^{S}L}
\end{array}\right..
\end{eqnarray}
Similar spinors are obtained when starting with the Hamiltonian Eq.
(\ref{eq:2L_Hamiltonian}) for the strained bilayer case, although they become 1$\times$8
arrays. We focus our attention on the strained monolayer
Josephson junctions in this appendix. Matching the wave functions of the
superconducting and normal segments at the two interfaces generates
the following matrix for reflection and transmission coefficients.
\begin{align}
\emph{M}=\left(
           \begin{array}{cc}
             \emph{M}_{11} & \emph{M}_{12} \\
             \emph{M}_{21} & \emph{M}_{22} \\
           \end{array}
         \right)
\end{align}
\begin{widetext}
\begin{align}
\nonumber\emph{M}_{11}=\left(
                \begin{array}{cccc}
                  -a_{e^{-}}^{S}e^{i\beta} & a_{h^{+}}^{S} & -a_{e^{+}}^{N} & a_{e^{-}}^{N} \\
                  e^{i\beta} & 1 & -1 & -1 \\
                  -a_{e^{-}}^{S}e^{-i\phi} & a_{h^{+}}^{S}e^{-i(\phi-\beta)} & 0 & 0 \\
                  e^{-i\phi} & e^{-i(\phi-\beta)} & 0 & 0 \\
                \end{array}
              \right)
\nonumber\emph{M}_{12}=\left(
                \begin{array}{cccc}
                  0 & 0 & 0 & 0 \\
                  0 & 0 & 0 & 0 \\
                  a_{h^{+}}^{N} & -a_{h^{-}}^{N} & 0 & 0 \\
                  -1 & -1 & 0 & 0 \\
                \end{array}
              \right)
\nonumber\emph{M}_{21}=\left(
                \begin{array}{cccc}
                  0 & 0 & -a_{e^{+}}^{N}e^{i\hbar k_{e}^{N} L} & a_{e^{-}}^{N}e^{-i\hbar k_{e}^{N} L} \\
                  0 & 0 & -e^{i\hbar k_{e}^{N}L} & -e^{-i\hbar k_{e}^{N}L} \\
                  0 & 0 & 0 & 0 \\
                  0 & 0 & 0 & 0 \\
                \end{array}
              \right)
\end{align}
\begin{align}
\nonumber\emph{M}_{22}=\left(
                \begin{array}{cccc}
                  0 & 0 & a_{e^{+}}^{S}e^{i(\hbar k_{e}^{S}L+\beta)} & -a_{h^{-}}^{S}e^{-i\hbar k_{h}^{S}L} \\
                  0 & 0 & e^{i(\hbar k_{e}^{S}L+\beta)} & e^{-i\hbar k_{h}^{S}L} \\
                  a_{h^{+}}^{N}e^{i\hbar k_{h}^{N}L} & -a_{h^{-}}^{N}e^{-i\hbar k_{h}^{N}L} & a_{e^{+}}^{S}e^{i\hbar k_{e}^{S}L} & -a_{h^{-}}^{S}e^{-i(\hbar k_{h}^{S}L-\beta)} \\
                  -e^{i\hbar k_{h}^{N}L} & -e^{-i\hbar k_{h}^{N}L} & e^{i\hbar k_{e}^{S}L} & e^{-i(\hbar k_{h}^{S}L-\beta)} \\
                \end{array}
              \right)
\end{align}
\end{widetext}
To determine the relation between the subgap energy of the quasiparticles and
the superconducting phase
difference, we use the determinant of $\emph{M}$ 
as mentioned in the Sec. \ref{section:theory}. Previously in this paper, we
have considered a heavily doped superconducting regions which dictates
normal trajectories of the quasiparticles relative the interfaces inside these
regions. In this appendix, we now allow for a moderately doped
superconducting region \ie $\mu^{S}>\varepsilon, \Delta$ and then
$\theta_{e}^{S}\approx\theta_{h}^{S}=\gamma\neq 0$. In this regime, 
we find $\digamma_1$, $\digamma_2$ and $\digamma_3$ factors as
follow;
\begin{widetext}
\begin{eqnarray}
&&\nonumber\digamma_1=\sin (k_{e}^{N}L) \sin (k_{h}^{N} L) (\sin
(\gamma )-\sin (\theta )) (\sin (\gamma )+\sin (\theta_A))-\cos
^2(\gamma ) \cos (\theta ) \cos (\theta_A) \cos (\phi )\\&&\nonumber
\digamma_2= \cos (k_{e}^{N}L)\cos (\gamma ) \cos (\theta ) \sin
(k_{h}^{N}L) (\sin (\gamma ) \sin (\theta_A)+1)\\&&\nonumber-  \cos
(k_{h}^{N}L)
   \cos (\gamma ) \cos (\theta_A) \sin (k_{e}^{N}L) (\sin (\gamma ) \sin (\theta
   )-1)\\&&\nonumber
   \digamma_3=\cos (k_{e}^{N}L) \cos (k_{h}^{N}L) \cos (\theta ) \cos
(\theta_A) \cos ^2(\gamma ) + (\sin (\gamma ) \sin (\theta )-1)
(\sin
   (\gamma ) \sin (\theta_A)+1)
\end{eqnarray}
\end{widetext}
We denote $\theta_e=\theta$ and $\theta_h=\theta_{A}$. If we apply
the short junction approximation to the factors and assume heavily
doped superconducting regions \ie $\gamma\rightarrow 0$, we recover the 
results of Ref. \onlinecite{cite:Titov1}.


\begin{thebibliography}{000}
\bibitem{cite:paper_sigma_i} We define $a_{x}=(1+ s)c_0 /2$, $a_{y}=c_0(1-
ps)/2$ so that the $Z$-strain leads
$\overrightarrow{\sigma_1}=\hat{x}a_{x}\sqrt{3}-\hat{y}a_{y}$,
$\overrightarrow{\sigma_2}=-\hat{x}a_{x}\sqrt{3}-\hat{y}a_{y}$,
$\overrightarrow{\sigma_3}=\hat{y}2a_{y}$ and then for the
$A$-strain,
$\overrightarrow{\sigma_1}=\hat{x}a_{y}\sqrt{3}-\hat{y}a_{x}$,
$\overrightarrow{\sigma_2}=-\hat{x}a_{y}\sqrt{3}-\hat{y}a_{x}$,
$\overrightarrow{\sigma_3}=\hat{y}2a_{x}$ . Also
$t_{i}=t_{0}e^{-3.37(\mid\overrightarrow{\sigma_{i}}\mid)/c_{0}-1}$,
$c_0=0.142 \AA$ is distance of two C atoms in the non-strained
graphene, $p=0.165$ is Poisson's ratio for graphite and $s=0.2$
which is fixed throughout our calculations, represents the strength
of applied tension to the system. The assumed tension strength is
near the gap threshold value, \ie $s=0.23\sim\%20$ (see
Ref.\onlinecite{cite:Pereira}).
\bibitem{cite:paper_delta_F} The evanescent modes appear in an interval
of order $\Delta E_F\simeq (\hbar v)^2/t_{\bot}L^2$
\cite{cite:Snyman}. Our results in the bilayer case are restricted
to the regime $\mu>\Delta$ due to the choice of wavefunction.
\end{thebibliography}
\end{document}